\documentclass{appolb}
\usepackage{latexsym}
\usepackage{graphicx}
\usepackage{bbm}
\usepackage{amsmath,amssymb, mathrsfs}
\usepackage[latin1]{inputenc}
\usepackage[dvips]{hyperref}
\usepackage{mciteplus}

\newcommand{\qs}{Q_\mathrm{s}}
\newcommand{\lqcd}{\Lambda_{_{\rm QCD}}}
\newcommand{\as}{{\alpha_\mathrm{s}}}
\newcommand{\et}{{\boldsymbol{e}}}

\newcommand{\xt}{{\boldsymbol{x}_\perp}}

\newcommand{\yt}{{\boldsymbol{y}_\perp}}
\newcommand{\ut}{{\boldsymbol{u}_\perp}}

\newcommand{\zt}{{\boldsymbol{z}_\perp}}
\newcommand{\pt}{{\boldsymbol{p}_\perp}}
\newcommand{\qt}{{\boldsymbol{q}_\perp}}
\newcommand{\kt}{{\boldsymbol{k}_\perp}}

\newcommand{\nabt}{\boldsymbol{\nabla}_\perp}

\newcommand{\gev}{\textrm{ GeV}}

\newcommand{\ra}{R_{_{\rm A}}}

\newcommand{\nr}[1]{(\ref{#1})} 
\newcommand{\ud}{\mathrm{d}}
\newcommand{\fig}{Fig.~}

\newcommand{\eq}{Eq.~}

\newcommand{\eqs}{Eqs.~}
\def\p{{\boldsymbol p}}
\def\q{{\boldsymbol q}}

\def\k{{\boldsymbol k}}

\def\x{{\boldsymbol x}}
\def\y{{\boldsymbol y}}

\def\v{{\boldsymbol v}}

\def\u{{\boldsymbol u}}

\begin{document}
\title{Initial conditions of heavy ion collisions and high energy factorization
\thanks{Presented at the XV Cracow Epiphany Conference
``Hadronic Interactions at the dawn of the
 LHC'', January 5-7, 2009. }
}
\author{T. Lappi
\address{Department of Physics
 P.O. Box 35, 40014 University of Jyv\"askyl\"a, Finland
\\
Institut de Physique Th\'eorique,
B\^at. 774, CEA/DSM/Saclay, \\  91191 Gif-sur-Yvette Cedex, France
}
}
\maketitle
\begin{abstract}
The ``Color Glass Condensate'' is an effective theory description for
 the small momentum fraction $x$ degrees of freedom in a high energy
 hadron or nucleus, which can be understood in terms of strong
 classical gluon fields. We discuss the resulting picture of the
 initial conditions in a relativistic heavy ion collision. We describe
 recent work to show that the leading logarithms of the collision
 energy can be factorized into the renormalization group evolution of
 the small $x$ wavefunction. We then describe how this framework can
 be used to understand the long range rapidity correlations observed
 by the RHIC experiments.
\end{abstract}
\PACS{13.85.Hd,24.85.+p,25.75.-q}

\section{The little bang of an ultrarelativistic heavy ion collision}

Quark gluon plasma is studied in the laboratory in collisions of heavy 
nuclei at ultrarelativistic energies, presently
 $\sqrt{s} = 200 A \gev$ at RHIC in Brookhaven or in the near future
$5500 A \gev$ at the LHC in CERN (where $A$ is the atomic number of the 
nucleus). The collision process is a complicated one, starting 
from the formation and equilibration of the matter to its evolution 
in time and space and ending in the 
decoupling of the system into 
the hadrons that are observed in the detectors.

The typical transverse momentum scales of the bulk of particles 
produced is in the GeV range, much less than the collision energy.
Thus the initial conditions depend on the 
small $x \sim p_T/\sqrt{s} \lesssim 0.01$ part of the nuclear
wavefunction. Because of the $\ln 1/x$ enhancement of soft gluon 
bremsstrahlung this is a dense gluonic system. When the occupation 
numbers of gluonic states in the wavefunction become large enough,
of the order of $1/\as$ (meaning that the gluon field
$A_\mu$ is of order $1/g$),
the nonlinear interaction part of the 
Yang-Mills Lagrangian becomes of the same order of magnitude as
the free part. The relevant comparison is between the
two terms in the covariant derivative
$D_\mu = \partial_\mu + i g A_\mu$: the momentum scales
$p_\mu = -i\partial_\mu \lesssim g A_\mu$ become nonlinear.
In the small $x$ wavefunction the relevant component is the transverse
momentum, we are therefore led to the concept of a transverse 
momentum scale $\qs$, the \emph{saturation scale}, below which the system is
dominated by nonlinear interactions. When the collision energy is high enough
($x$ small enough), $\qs \gg \lqcd$ and the coupling is weak: we are
faced with a \emph{nonperturbative} strongly interacting system
with a \emph{weak coupling constant}. On the other hand, the large occupation 
numbers mean that the system should behave as a \emph{classical} field.
 This suggests a way of organizing
calculations that differs from traditional perturbation theory. Instead 
of developing as a series of powers in $g A_\mu$ we want to calculate the
classical background field $A^\mu_{\textrm{cl.}}$ and loop corrections 
(which are suppressed by powers of $g$) to all orders in 
$g A^\mu_{\textrm{cl.}}$. The classical gluon field will then be radiated
by the large $x$ degrees of freedom, which we shall treat as 
effective classical color charges. This picture of the 
high energy wavefunction is referred to as the 
Color Glass condensate (CGC, for reviews see 
e.g.~\cite{Iancu:2003xm,*Weigert:2005us}). The collision of two such systems
leads, in the early stages $1/\sqrt{s} \ll \tau \lesssim 1/\qs$, 
to classical field configurations known 
as the Glasma~\cite{Lappi:2006fp,*Lappi:2006nx}.

At early times ($\tau \ll \ra$, see \fig\ref{fig:spacet} for the coordinate
system) the bulk of the system cannot, by causality, be aware of its finite
size in the transverse plane. It will therefore be in a longitudinally expanding,
to a first approximation boost invariant ($\partial_\eta = 0$) state,
a 1-dimensional Hubble expansion.
Boost invariance can come in two flavors. As we shall argue in the following, the
very early time glasma degrees of freedom are boost invariant at the level of 
\emph{field configurations}. This means that the longitudinal momenta of particles
redshift towards zero $p_z \sim 1/\tau$ 
while $p_T \sim$~constant and the system becomes very anisotropic in momentum space.
This field level invariance is broken by quantum fluctuations suppressed by $\as$,
 which then eventually
evolve into a more equilibrated fluid that is isotropic its local rest frame.
What remains is a boost invariant profile of particle flow, as in the Bjorken
hydrodynamical picture. What concerns us in this paper is the very earliest glasma
stage and the initial quantum fluctuations that serve as the seeds of isotropization.

In the following we shall first discuss the leading order,
 classical field level, results for the structure of the glasma fields
and gluon production. We shall then, in Sec.~\ref{sec:fact},
describe some ingredients of
the recent proof \cite{Gelis:2008rw} that shows how the leading logarithmic
divergences of the NLO corrections to these fields can be absorbed 
into the renormalization group (RG) evolution of the weight functionals
describing the hard sources, the 
\emph{JIMWLK factorization theorem}. In Sec.~\ref{sec:multig} we shall 
then describe multigluon correlations in the same framework.

\section{Gluon production to leading order and the glasma}

\begin{figure}
\centerline{\includegraphics[width=0.7\textwidth]{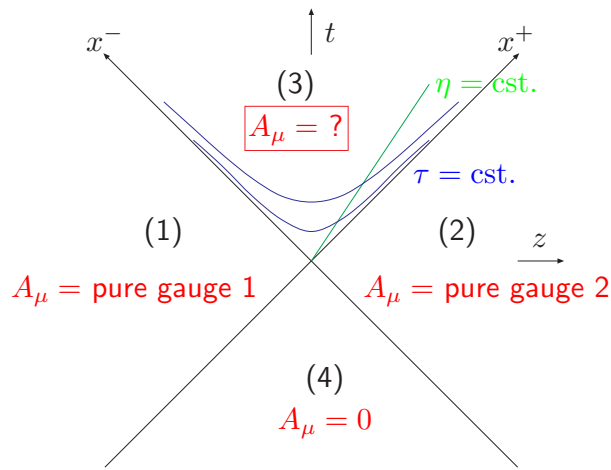}}
\caption{Spacetime structure of the CGC and  glasma fields.
It is convenient to use the  
coordinate system with proper time $\tau =\sqrt{2x^-x^+}$ and
the spacetime rapidity $\eta = \half \ln{x^+/x^-}$.}
\label{fig:spacet}
\end{figure}

The CGC framework is based on a separation of
scales between small $x$ and large $x$ degrees of freedom, which are 
treated as a classical field and an effective color charge density.
In practice the classical field is obtained from the equation of motion
\begin{equation}\label{eq:ym}
[D_{\mu},F^{\mu \nu}] \quad = \quad J^{\nu}.
\end{equation}
The current in the case of a nucleus-nucleus consists
of two  infinitely Lorentz-contracted (this picture 
will be discussed more below) nuclei on 
the light cone~\cite{Kovner:1995ts,*Kovchegov:1997ke}:
\begin{equation}\label{eq:source}
J^{\mu} =\delta^{\mu +}\rho_{(1)}(\xt)\delta(x^-) 
+ \delta^{\mu -}\rho_{(2)}(\xt)\delta(x^+).
\end{equation}
The large $x$ degrees of freedom have now been reduced to a 
classical effective color charge density $\rho(\xt)$, which is
a static (hence the ``glass'') stochastic variable. Its values
are drawn from a  probability distribution
$W_y[\rho(\xt)]$ which depends on the cutoff rapidity $y=\ln 1/x$
separating large and small $x$.
To a first approximation we can take e.g. the Gaussian distribution
of color charges that defines the 
MV~\cite{McLerran:1994ni,*McLerran:1994ka,*McLerran:1994vd}
model
\begin{equation}
W[\rho(\xt)] = \mathcal{N} \exp \left[
 - \half \int \ud^2\xt \rho^a(\xt)\rho^a(\xt)/g^2\mu^2
\right].
\end{equation}
The probability distribution $W_y[\rho(\xt)]$ is analogous to a parton 
distribution function in the DGLAP formalism; it is a nonperturbative input
that we are not able to compute from first principles, but one can derive 
evolution equation for its $y$--dependence.
This equation is known by the acronym JIMWLK.

\begin{figure}
\centerline{
\includegraphics[width=0.45\textwidth]{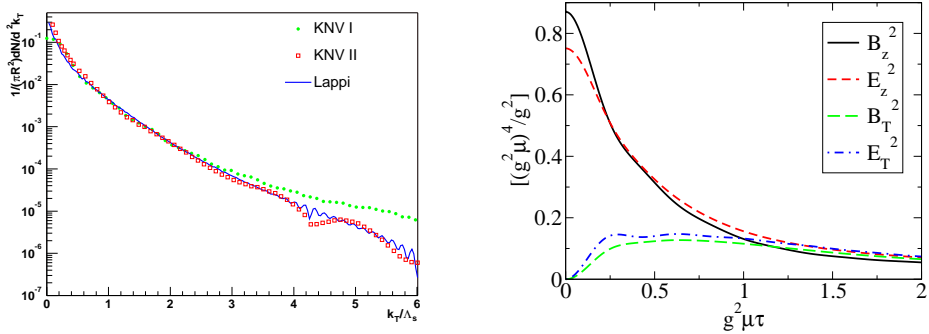}
\hfill
\includegraphics[width=0.45\textwidth]{components.eps}
}
\caption{
Left: Gluon spectrum from the leading order classical field computation.
Right: the components of the glasma field, the initial condition is a longitudinal 
electric and magnetic field, the transverse components develop in a time
$\sim 1/\qs$.}
\label{fig:spect}
\end{figure}

For a fixed configuration of the color sources $\rho$ the calculation of the 
Glasma fields proceeds as follows~ \cite{Kovner:1995ts}.
The solution of the Yang-Mills equations in the regions of spacetime
$x^\pm>0,x^\mp<0$ that 
are causally connected to only one of the nuclei
(areas (1) and (2) in \fig\ref{fig:spacet}) 
is an analytically known pure gauge field. It 
gives the initial condition for the numerical solution in the
forward light cone (3). Working in the temporal gauge
$A_\tau=0$ these initial conditions are 
\begin{eqnarray}\label{eq:trinitcond}
A^i|_{\tau=0} &=& A^i_{(1)} + A^i_{(2)} \\
\label{eq:longinitcond}
A^\eta|_{\tau=0} &=& \frac{ig}{2}[A^i_{(1)},A^i_{(2)}],
\end{eqnarray}
where $A^i_{(1,2)}$ are the pure gauge fields that are the solutions
of the one-nucleus problem
\begin{equation} \label{eq:LCsol}
A^i_{(1,2)}  =  \frac{i}{g} U_{(1,2)}(\xt) \partial_i U_{(1,2)}^\dag(\xt).
\end{equation}
These pure gauge fields are gauge transforms of the vacuum with 
the \emph{Wilson lines} computed from the color charge density
\begin{equation} \label{eq:pathorder}
U_{(1)}(\xt,x^-) = \mathrm{P} \exp \left\{ - ig \int_{-\infty}^{x^-} 
\!\!\!\ud y^-
\frac{\rho(\xt,y^-)}{  \nabt^2}
\right\},
\end{equation}
with the Wilson line $U_{(2)}$ given by the analogous formula in terms of the 
other color charge density.

These Wilson lines are in fact the most natural variables to describe 
the soft gluonic field degrees of freedom of the nucleus; they correspond
to the eikonal scattering amplitude of a color charge off the strong color fields.
For example the dipole cross that determines the structure function measured in 
deep inelastic scattering is a correlator of these same Wilson lines.
The upper limit of the $y^-$--integral in \eq\nr{eq:pathorder} must be thought 
of as $x^- \sim e^{y}$; when the cutoff rapidity $y$ becomes larger ($x$ smaller),
smaller momentum $p^+$ gluons are considered as part of the source, 
which consequently extends further in the conjugate variable $x^-$. Thus each
infinitesimal step 
in the renormalization group evolution towards smaller $x$ corresponds to
adding a layer in $x^-$ to the color source, or equivalently to 
multiplying the Wilson line by an SU(3) matrix that is infinitesimally 
close to identity.

From the point of view of the classical glasma fields
in \eq\nr{eq:LCsol}
the Wilson lines are independent of the longitudinal coordinate: the 
longitudinal structure appears only indirectly in the properties
of the probability distribution $W_y[U]$. The classical fields represent 
degrees of freedom with a smaller $p^+$ than the ones integrated out to the
Wilson lines and are not able to resolve their structure which is shorter 
range in $x^-$. This is the sense in which the
$\delta$-functions in the currents of \eq\nr{eq:source} must be understood.

The numerical method for solving the Yang-Mills equations in the forward 
light cone was developed in Ref.~\cite{Krasnitz:1998ns}
and the actual computations reported 
in Ref.~~\cite{Krasnitz:1999wc,*Krasnitz:2000gz,*Krasnitz:2001qu,*Krasnitz:2002mn,%
*Krasnitz:2003jw,*Lappi:2003bi}
The equations of motion are most conveniently solved 
 in the Hamiltonian formalism. Due to the
boost invariance of the initial conditions  in the high energy limit
the Yang-Mills equations
can be dimensionally reduced to a 2+1 dimensional gauge theory with
the $\eta$--component of the gauge field becoming an adjoint
scalar field. With the assumption of boost invariance one is explicitly
neglecting the longitudinal momenta of the gluons. In the Hamiltonian 
formalism one obtains directly the (transverse) energy. By decomposing
the fields in Fourier modes one can also define a gluon multiplicity
corresponding to the classical gauge fields; the resulting gluon spectrum
is shown in \fig\ref{fig:spect}.
The color fields of the two nuclei are transverse electric and magnetic 
fields on the light cone. 
The glasma fields left over in the region between the two nuclei 
after the collision at times $1 \leq \tau \leq 1/\qs$
are, however, longitudinal along the beam axis~\cite{Lappi:2006fp} 
(see \fig\ref{fig:spect}).

\section{Factorization}
\label{sec:fact}

\begin{figure}
\centerline{\includegraphics[width=0.7\textwidth]{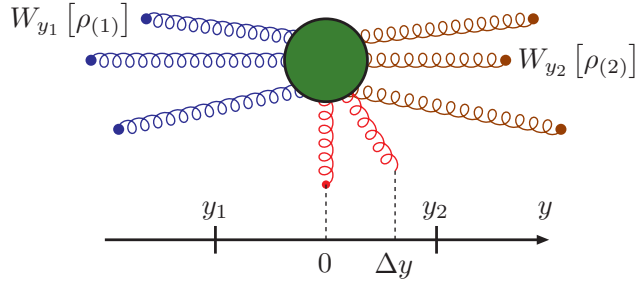}}
\caption{
Factorization of LLog corrections to gluon production: the phase space 
integral over $\Delta y$ diverges and is cut off at the separation scales
$y_{1,2}$. The dependence of the color charge density distributions
$W_{y_{1,2}}$ on the cutoff cancels the leading logarithmic part of 
the dependence on $y_{1,2}$.
}
\label{fig:facth}
\end{figure}

To understand the context of the high energy factorization 
theorem proven in Ref.~\cite{Gelis:2008rw} it
is perhaps useful to look first at the weak field
limit of the CGC, where particle production can be computed using  $k_T$-factorization
(\cite{Gribov:1984tu}, see e.g.~\cite{Kharzeev:2003wz}
for an application to heavy ion collisions).
The leading order multiplicity  is
\begin{equation}
\frac{\ud N}{\ud^2\pt \ud y} =
\frac{1}{\as}\frac{1}{\pt^{\!\!\!2}} \int
 \frac{\ud^2\kt}{(2 \pi)^2} 
\varphi_y(\kt) \varphi_y(\pt - \kt).
\end{equation}
To obtain  the real part of the leading log correction to this result one must take the
corresponding expression for double inclusive gluon production
\begin{equation}
\frac{\ud N}{\ud^2\pt \ud y_p \ud^2\qt \ud y_q} =
\frac{1}{\as}\frac{1}{\pt^{\!\!\!2} \qt^{\!\!\!2}} \int \frac{\ud^2\kt}{(2 \pi)^2}
\varphi_y(\kt_1) \varphi_y(\pt + \qt - \kt).
\end{equation}
and integrate it over the phase space of the second gluon $(\qt,y_q)$. 
Note that at leading log accuracy we have here taken
the multi-Regge kinematical limit, assuming
that the two produced gluons are far apart in rapidity (see 
e.g.~\cite{Leonidov:1999nc}).
The integral over $y_q$ diverges linearly (this is the general behavior of 
the $gg \to gg$ scattering 
amplitude in the high energy limit $t$ fixed, $s\sim -u \to \infty$). 
This divergence is compensated (to the appropriate order in $\as$) by the real part of 
the BFKL evolution equation for $\varphi_y(\kt_1)$.

In the fully nonlinear case of AA collisions the $k_T$-factorization is broken
(see e.g.~\cite{Krasnitz:1998ns,Blaizot:2008yb}), and
one must solve the equations of motion to all orders in the strong classical field.
The analogue of the unintegrated parton distribution $\varphi_y(\kt)$ is
the color charge density distribution $W_y[\rho]$. These are similar in the
sense that they are not (complex) wavefunctions but (at least loosely speaking)
real probability distributions. Factorization can be understood as a statement that 
one has found a convenient set of degrees of freedom 
in which one can compute physical observable from only the diagonal
elements of the density matrix of the incoming nuclei.
The difference is that when in the dilute case these degrees of freedom are
 numbers of  gluons with a given momentum, in the nonlinear case the
appropriate variable  is the color charge
density and the relevant evolution equation is JIMWLK, not BFKL.
The kinematical situation, however, remains the same. To produce a gluon at a very large 
rapidity (or a contribution in the loop integral of the virtual contribution with a large
$k^+$) one must get a large $+$-momentum from the right-moving source. 
Thus one is probing the  source at a large $k^+$, i.e. small distances in
$x^-$, and the result must involve $W_y[\rho]$ at a larger rapidity
(see Fig.~\ref{fig:facth}).

The underlying physical reason for factorization is that this
fluctuation with a large $k^+$ requires such a long interval in $x^+$ 
to radiated that it must be produced well before and independently of the
interaction with the other 
(left moving and thus localized in $x^+$) source.
 The concrete task is then to show that when one computes the NLO corrections
to a given observable in the Glasma, all the leading 
logarithmic divergences can be absorbed into the RG evolution of the sources
with the same Hamiltonian that was derived by considering only the DIS process. 
This is the  proof~\cite{Gelis:2008rw,Gelis:2008ad,Gelis:2007kn} of factorization that 
we will briefly describe in the following.

Consider the single inclusive gluon multiplicity
which is a sum of probabilities to produce $n+1$ particles, with the phase space
of the additional $n$ must be integrated out
\begin{equation}
\frac{\ud N}{\ud^3\vec\p}\sim \sum_{n=0}^\infty 
\frac{1}{n!}\int\left[\ud^3\vec\p_1\cdots \ud^3\vec\p_n\right]\;
\left|\left<{\vec\p}\;\,\vec\p_1\cdots\vec\p_n\big|0\right>\right|^2.
\end{equation}

\begin{figure}
\begin{center}
\includegraphics[width=0.35\textwidth]{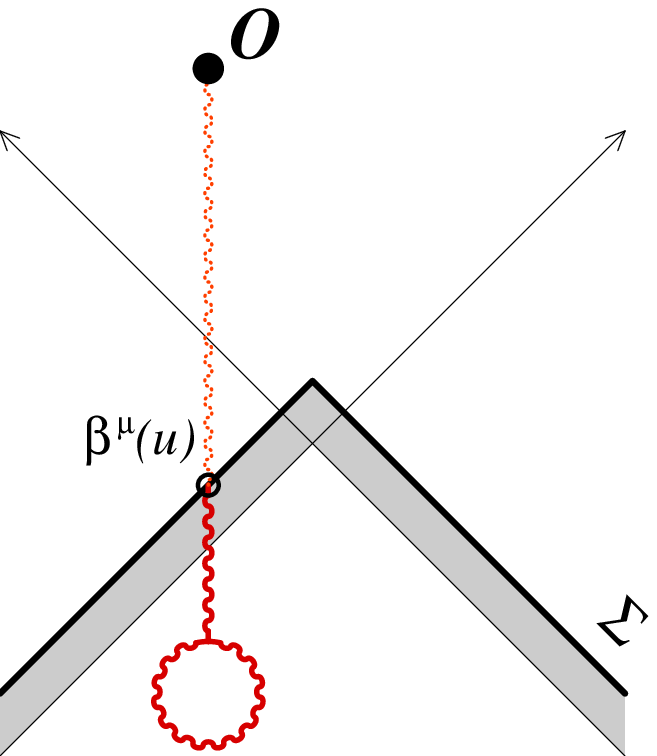}
\includegraphics[width=0.35\textwidth]{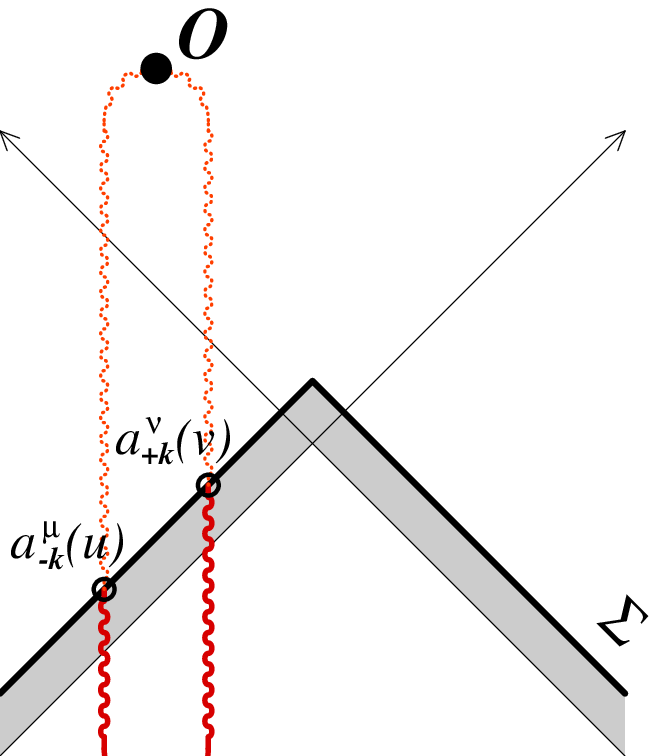}
\end{center}
\caption{The one loop one and two point functions in the background field, separated into
the parts before the light cone $\Sigma$.}
\label{fig:betagpm}
\end{figure}

Because we have a theory with external color sources of order $\rho \sim 1/g,$
all insertions of the sources appear at the same order in $g$~\cite{Gelis:2006yv}. 
A calculation using the 
Schwinger-Keldysh formalism leads to the following results:
At LO, the multiplicity is obtained from the  \emph{retarded} solution of classical 
field equations (here $(\dots)$ includes the appropriate normalization and projection
to physical polarizations)
\begin{equation}
\frac{\ud N_{_{\rm LO}}}{\ud^3\vec\p}
=
\int  \ud^3 \x \ud^3 \y
e^{i\vec\p\cdot(\vec\x-\vec\y)}\;
(\cdots)
\left[
{{\cal A}^\mu(t,\vec\x)}{{\cal A}^\nu(t,\vec\y)}
\right]
\Big|_{t\to\infty}.
\end{equation}
The NLO contribution includes the one loop correction to the classical field and the 
$+-$ component of the Schwinger-Keldysh (SK) propagator in the background field
\begin{multline}
\frac{\ud N_{_{\rm NLO}}}{\ud^3\vec\p}
=
\int \ud^3 \x \ud^3 \y
e^{i\vec\p\cdot(\vec\x-\vec\y)}\;
(\cdots)
\Big[
{\mathcal{G}_{+-}^{\mu\nu}(x,y)}
+
\\
{\beta_+^\mu(t,\vec\x)}\;{\mathcal{A}_-^\nu(t,\vec\y)}
+
{\mathcal{A}_+^\mu(t,\vec\x)}\;{\beta_-^\nu(t,\vec\y)}
\Big]
\bigg|_{t\to\infty}.
\end{multline}
Now consider a small fluctuation $a^\mu(x)$ of the gluon field around the classical
value. The $+-$ (SK index) component of the propagator is bilinear in 
these small fluctuations satisfying \emph{retarded} boundary conditions.
Also the virtual term $\beta$ satisfies an equation of motion with
 a retarded  boundary condition and a source term involving a loop in the 
classical background field,
see Fig.~\ref{fig:betagpm} for a pictorial representation of this structure.
One can express the propagation of such a small
fluctuation $a^\mu(x)$ above the  past light cone $\Sigma$ as 
a functional derivative $\mathbb{T}_{\u}$ of the LO classical field $\mathcal{A}^\mu(x)$ 
with respect to its  initial condition on $\Sigma$:
$a^\mu(x) =\int_{\vec\u\in{\rm \Sigma}} a(\vec\u) \cdot
 \mathbb{T}_{\u} \mathcal{A}^\mu(x).$
This leads after some rearrangements to the expression for the NLO contribution to the multiplicity
as a functional derivative operator acting on the leading order result:
\begin{equation} \label{eq:dop}
\left.\frac{\ud N}{\ud^3\vec\p}\right|_{_{\rm NLO}}
=
\left[
\frac{1}{2} \! \int_\Sigma \! \ud^3\u \ud^3\v
\mathcal{G}_{\mu \nu}(\u,\v) \mathbb{T}^\mu_{\u} \mathbb{T}^\nu_{\v}
+\int_\Sigma \! \ud^3\u
\mathbf{\beta}_\mu(\u) \mathbb{T}^\nu_{\u}
\right]
\left.\frac{\ud N}{\ud^3\vec\p}\right|_{_{\rm LO}}. 
\end{equation}
This expression involves the part of the two point function 
below the light cone $\Sigma$:
\begin{equation}
\mathcal{G}^{\mu\nu}(\vec\u,\vec\v)\equiv
\int\frac{\ud^3\vec\k}{(2\pi)^3 2E_\k}\; a^\mu_{-\k}(\u)\,a^\nu_{+\k}(\v).
\end{equation}
Here the small fluctuation field $a^\mu(x)$ is the solution of the linearized
equation of motion in the classical field background with an 
initial condition given by a plane wave
$\lim_{x^0\to -\infty}a^\mu_{\pm \k}(x) = \epsilon^\mu(\k)e^{\pm i k\cdot x}$.

The leading logarithmic contribution comes from the longitudinal component of the 
integral over $\k$, the momentum of the initial plane wave perturbation (and the 
corresponding momentum in the one loop source term for the equation of motion
satisfied by $\beta$). This LLog part of the functional derivative \nr{eq:dop} operator
turns out to be precisely equivalent to the sum of the JIMWLK Hamiltonians
describing the RG evolution of the source distributions $W_y[\rho]$.
 The JIMWLK  Hamiltonian 
\begin{equation}\label{eq:hjimwlk}
\mathcal{H}
\equiv \frac{1}{2}
\int \ud^2\xt \ud^2\yt 
D_a(\xt)
\eta^{ab}(\xt,\yt)
D_b(\yt)
\end{equation}
is most naturally expressed in terms of Lie derivatives $D_a(\xt)$ operating on
 the Wilson lines introduced in \eq\nr{eq:pathorder}.
in terms of which the kernel in \eq\nr{eq:hjimwlk} is
\begin{multline}
\eta^{ab}
(\xt,\yt)
=
\frac{1}{\pi}
\int \ud^2\ut
\frac{(\xt-\ut)\cdot(\yt-\ut)}{(\xt-\ut)^2(\yt-\ut)^2}\;
\Big[
U(\xt) U^\dag(\yt)
\\
-U(\xt) U^\dagger(\ut)
-U(\ut) U^\dag(\yt)
+1\Big]^{ab} 
\label{eq:eta-f}.
\end{multline}
The fact that no other terms with the same logarithmic divergences appear is the 
proof of factorization; this is the central result of Ref.~\cite{Gelis:2008rw}.

\section{Multigluon production}
\label{sec:multig}

\subsection{Short range in rapidity}

Let us then consider the probability distribution of the number of gluons
produced in a small rapidity interval. It was shown in Ref.~\cite{Gelis:2008ad}
that a similar factorization theorem holds for the leading logarithmic
corrections to this probability distribution in the sense that we will briefly review
here. 
It is convenient to define a generating functional
\begin{equation}\label{eq:moments}
\mathcal{F}[z(\p)]=\sum_{n=0}^{\infty}\frac{1}{n!}
\int \left[
\prod_{i=1}^n
\ud^3\p_i\;
(z(\p_i)-1)\right]
\frac{\ud^n N_n}{\ud^3\p_1\cdots \ud^3\p_n}.
\end{equation}
The Taylor coefficients of $\mathcal{F}$ around $z=1$ correspond to the
moments of the probability distribution; integrated over the momenta of 
the produced gluons they are
\begin{equation}
\left\langle N \right\rangle \quad 
 \left\langle N(N-1) \right\rangle  \quad \dots 
\quad \left\langle N(N-1)\cdots(N-n+1) \right\rangle.
\end{equation}
The result of  Ref.~\cite{Gelis:2008ad} is that when these moments
are calculated to NLO accuracy, the leading logarithms 
can be resummed into the JIMWLK evolution of the sources 
completely analogously to the single inclusive gluon distribution.
The resulting probability distribution can be written as:
\begin{equation}\label{eq:probadist}
\frac{\ud^n P_n}{\ud^3\p_1\cdots \ud^3\p_n}
=
\int\limits_{\rho_1,\rho_2}
W_{Y}\big[\rho_1\big]  W_{Y}\big[\rho_2\big]
  \frac{1}{n!}  \frac{\ud N}{\ud^3\p_1}
\cdots  \frac{\ud N}{\ud^3\p_n}
e^{ -\int \ud^3\p \frac{\ud N}{\ud^3\p}}.
\end{equation}
Note that  the
Poissonian-looking form of the result is to some extent an artifact of our choosing
to develop and truncate precisely the moments \eq\nr{eq:moments} that
are simply $\langle N \rangle^n$ for a Poissonian distribution. Since in our
power counting $N\sim 1/\as$, any contributions that would make the
distribution \eq\nr{eq:probadist} deviate from the functional form are
of higher order in the weak coupling expansion of the moments \nr{eq:moments} 
and are neglected in our calculation unless they are enhanced by large 
logarithms of $x$. Nevertheless it should be emphasized that in spite of appearances
of \eq\nr{eq:probadist} the probability distribution is in fact not
Poissonian. To understand the nontrivial nature of this result 
it must be remembered that the individual factors
of $\frac{\ud N}{\ud^3\p_i}$ in \eq\nr{eq:probadist} are all functionals
of the \emph{same} color charge densities $\rho_{1,2}$; thus the averaging 
over the $\rho$'s induces a correlation between them. These correlations 
are precisely the leading logarithmic modifications to the probability distribution;
they have been resummed into the distributions $W_y$; the functional 
form of the multigluon correlation function \emph{under} the functional 
integral in \eq\nr{eq:probadist} is the same as at leading order. This is the
result of the proof in Ref.~\cite{Gelis:2008ad}.

\subsection{Long range in rapidity}

Let us now relax the restriction that the gluons should be observed only
in a small rapidity interval and allow for arbitrary separations in 
rapidity~\cite{Gelis:2008sz}. There is now phase space available to radiate
gluons (and for the corresponding virtual contributions) between the 
measured gluons, and including this radiation can introduce additional
large logarithms of the energy.
To develop a physical picture of this situation it is perhaps useful to 
take a step back and consider a more general picture of JIMWLK
evolution  in terms of its Langevin formulation  
derived in Ref.~\cite{Blaizot:2002xy}. The original derivation is 
presented purely as an alternative formulation to generate the single Wilson
line probability distribution that solves the JIMWLK equation. 
The JIMWLK equation as it is usually written, as an equation satisfied
by the probability distribution of Wilson lines at a single
rapidity $y$, does not formally give
information about correlations between different rapidities.
Going back to the derivation one sees, however, that the rapidity correlations
are also encoded in the formalism. This is most transparent in the
Langevin formulation, where one can identify each
trajectory in the Langevin equation with one high energy collision
event. In this sense the Langevin formulation contains more physical
information than just the JIMWLK equation for the probability
distribution at a single rapidity; it also gives the combined
probability distribution for Wilson lines at different rapidities
\begin{equation}\label{eq:npointdiscr}
W_{y_1 \dots y_n} [U_1(\xt), \dots, U_n(\xt)].
\end{equation}
Knowing the general (multiple rapidity) probability distribution
will enable us to compute the correlations between Wilson lines,
and consequently of physical observables such as the multiplicities,
at different rapidities. In the following we will give a more precise
formulation of this statement and show that it is consistent with our
previous result concerning multigluon production.

In the Langevin formulation the distribution $W$ of the Wilson lines
can be obtained by evolving in rapidity the elements of an ensemble of
Wilson lines according to
\begin{equation}
U(y + \ud y,\xt) = U(y,\xt)e^{-i \alpha(y,\xt) \ud y}
\label{eq:langevin}
\end{equation}
where the change in a small step $\ud y$ in rapidity is given by a
deterministic term and a stochastic term,
\begin{equation}
\alpha^a(y,\xt) = \sigma^a(\xt,y) + \int_{\zt} 
\et^{ab}(\xt,\zt) \zeta^b(\zt,y).
\end{equation}
Here $\zeta^b$ is a Gaussian random variable defined by $\langle
\zeta_i^a(\xt,y) \zeta_j^b(\yt,y') \rangle = \delta^{ab}\delta_{ij}
\delta^2(\xt-\yt) \delta(y-y')$
and the square root of the JIMWLK kernel is
\begin{equation}\label{eq:sqrteta}
\et^{ac}(\xt,\zt) \equiv \frac{1}{\sqrt{4 \pi^3}}
\frac{\xt - \zt}{(\xt - \zt)^2} 
(1\!\!-\!U^\dag(\xt) U(\zt))^{ac}.
\end{equation}
This stochastic
formulation is the method used in numerical studies of the JIMWLK
equation~\cite{Rummukainen:2003ns}. 

Knowing that the correlation follows from a Langevin equation 
imposes an additional structure
(of a Markovian process) on the probability distribution:
\begin{equation} \label{eq:markov}
W_{y_p,y_q}\left[U^p,U^q\right] 
=
G_{y_p-y_q}\left[U^p,U^q\right] 
W_{y_q}\left[U^q\right],
\end{equation}
where the JIMWLK Green's function $G$ is determined by the initial condition
\begin{equation} \label{eq:greenic}
\lim_{y_p \to y_q} G_{y_p-y_q}\left[U^p,U^q\right] 
=
\delta\left(U^p(\xt)-U^q(\xt)\right)
\end{equation}
and the requirement that it must satisfy the JIMWLK equation
\begin{equation}
\partial_{y_p} G_{y_p-y_q}\left[U^p,U^q\right] 
= \mathcal{H}\left(U^p(\xt)\right)G_{y_p-y_q}\left[U^p,U^q\right].
\end{equation}
This JIMWLK Green's function contains all the information, at the leading 
log level, of long range rapidity correlations in gluon production. 
This structure follows from the
computation of the leading log part of 1-loop corrections  to  a 
wide class of observables that can be expressed in terms of correlators
of the gluon fields at $\tau=0$ (or equivalently Wilson lines)

\begin{equation}
\left<{\cal O}\right>_{_{\rm LLog}}
=
\int 
\left[DU_1(y,\xt)\right]\left[DU_2(y,\xt)\right]
W\left[U_1(y,\xt)\right]W\left[U_2(y,\xt)\right]
\mathcal{O}_{_{\rm LO}}.
\label{eq:fact-gen}
\end{equation}
Here we have introduced a continuous rapidity notation 
$W[U(y,\xt)] $for the 
probability distribution of Wilson lines \nr{eq:npointdiscr}.
This should be understood as a probability distribution
for the \emph{trajectories} that the Wilson line
$U(\xt)$ takes on the group manifold along its evolution
forward in $y$ following the Langevin equation. We can formally 
return from the distribution of trajectories to a distribution 
of Wilson lines at one individual rapidity as
\begin{equation}
W_{y}[U(\xt)]
\equiv
\int\left[DU(y,\xt)\right]\;
W\left[U(y,\xt)\right]
\delta\left[U(\xt)-U(y,\xt)\right].
\end{equation}

Equation \nr{eq:fact-gen} is the central result of Ref.~\cite{Gelis:2008sz}, 
showing that all the leading logarithms of rapidity (either the rapidity
intervals between the nuclei and the tagged gluons, or between the
various produced gluons) can be absorbed into the probability
distributions $W$ for the trajectories of Wilson lines of the two
projectiles.
However, the crucial point to keep in mind is that it
involves an average over $y$-dependent ``trajectories'' of Wilson
lines, rather than an average over Wilson lines at a given fixed
rapidity.

\begin{figure}
\centerline{\includegraphics[width=0.7\textwidth]{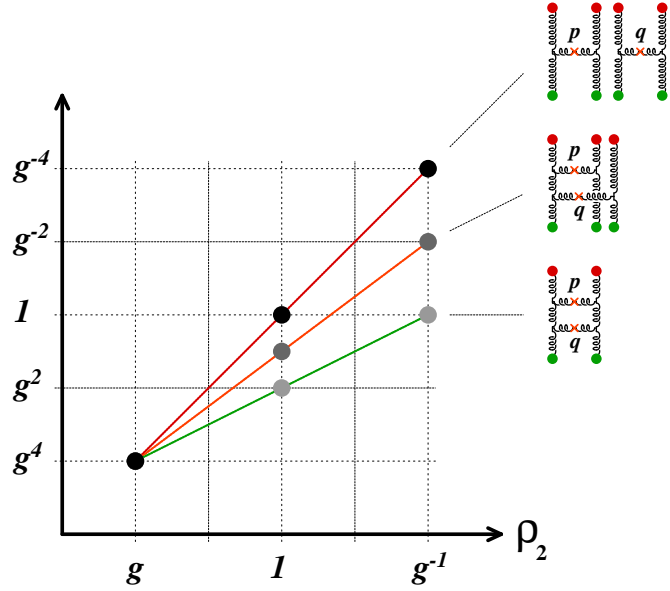}}
\caption{Relative importance of connected and disconnected diagrams to the
two gluon correlation function. One of the color charge densities is 
considered large, $\rho_1 \sim 1/g$, whereas the  other is allowed to vary between the
``AA'' case $\rho_2 \sim 1/g$ and the ``pA'' one $\rho_2 \sim g$. The order of the
disconnected diagram, on top, is $g^4 \rho_1^4 \rho_2^4$, whereas the
 interference diagram in the middle is $g^4 \rho_1^3 \rho_2^3$ and the
 connected one, lowest, is $g^4 \rho_1^2 \rho_2^2$.
In the ``AA'' case the disconnected diagram dominates, for the ``pA'' case all three
are equally important. In the dilute ``pp'' limit only the connected diagram
matters and both gluons are produced from the same BFKL ladder.
}\label{fig:diags}
\end{figure}

The calculation of multigluon correlation is in fact 
simplified in the  strong field limit, where the leading contribution to particle
production corresponds to the classical field and the correlations 
are encoded in the evolution of the sources~\cite{Kharzeev:2004bw,*Armesto:2006bv}.
In the ``pA'' case where one of the sources is assumed to be dilute, the situation
becomes much more complicated, because the disconnected classical contributions
are not the only dominant ones any more. This structure is illustrated in 
\fig\ref{fig:diags}.

\subsection{Application to multigluon correlations}

Let us now specialize \eq~(\ref{eq:fact-gen}) to the case of the
single and double inclusive gluon spectra. The single
inclusive gluon spectrum $\ud N_1/\ud^3\p$ at LO depends only on
Wilson lines $U_{1,2}(y_p,\xt)$ at the rapidity $y_p$ of the
produced gluon.
One then obtains the known result for
the single inclusive gluon spectrum as
\begin{equation}
\left.
\frac{\ud N_1}{\ud^2\p_\perp \ud y}
\right|_{_{\rm LLog}}
\!\!\!
=
\!
\int
\left[D U_{1}\right]\left[ DU_{2}\right] 
W_{y_p}\left[U_{1}\right]\,
W_{y_p}\left[U_{2}\right]\;
\left.
\frac{\ud N_1\big[U_{1},U_{2}\big]}{\ud^2\p_\perp \ud y}
\right|_{_{\rm LO}}.
\label{eq:N1-resummed}
\end{equation}
For the resummed inclusive two-gluon spectrum, we must recall that at
LO it is simply the product of two single gluon spectra, each of which
depends on Wilson lines at the rapidity of the corresponding gluon. It
is then straightforward to proceed as in the case of the single gluon
spectrum in order to obtain~:
\begin{multline}
\left.
\frac{\ud N_2}{\ud^2 \pt \ud y_p  \ud^2 \qt \ud y_q}
\right|_{_{\rm LLog}}
\!\!\!
=
\!
\int
\left[D U^p_{1}\right]\left[D U^p_{2}\right]\left[D U^q_{1}
\right]\left[D U^q_{2}\right] 
\times
\\
W_{y_p,y_q}\left[U^p_{1},U^q_{1}\right]\,
W_{y_p,y_q}\left[U^p_{2},U^q_{2}\right]
\left.
\frac{\ud N_1\left[U_{1}^p,U_{2}^p\right]}{\ud^2\p_\perp \ud y_p}
\right|_{_{\rm LO}}
\left.
\frac{\ud N_1\left[U^q_{1},U^q_{2}\right]}{\ud^2\q_\perp \ud y_q}
\right|_{_{\rm LO}},
\label{eq:N2-resummed-largey}
\end{multline}
where the double probability distribution $W_{y_p,y_q}\left[U^p_{1},U^q_{1}\right]$ 
is given by \eqs\nr{eq:markov} and~\nr{eq:greenic}.

We have now assembled all the 
ingredients needed to compute rapidity correlations in the Glasma and address
features such as the elongated ``ridge'' structure in the two particle
correlation observed at RHIC~\cite{Daugherity:2008su,*Putschke:2007mi,*Wenger:2008ts}.
 There have already been several boost invariant
classical field calculations~\cite{Dumitru:2008wn,*Gavin:2008ev} 
of this effect and the azimuthal structure,
 but the inclusion 
of quantum evolution is needed to understand the rapidity dependence.
As a first approximation one should be able to
formulate the equivalent of the mean field approximation leading to the BK
equation for the JIMWLK propagator. Numerical studies of the JIMWLK equation
would then be needed to study the validity of this approximation
for rapidity correlations; in the structure of the single rapidity probability 
distribution the violations from the mean field limit have been observed to be 
small~\cite{Kovchegov:2008mk}.

\subsection*{Acknowledgments}
I am grateful to the organizers for their invitation to this meeting dedicated
to the memory of Jan Kwieci\'nski.
This talk is based on work done in collaboration with F. Gelis 
and R. Venugopalan.
The author is supported by the Academy of Finland, contract 126604.

\bibliographystyle{h-physrev4mod2M}
\bibliography{spires}

\end{document}